\documentclass[fleqn,10p]{wlscirep}
\usepackage{makecell}
\usepackage{caption}

\title{Size effects in micro and nanoscale materials fracture.}

\author[1]{Alessandro Taloni}
\author[2]{Michele Vodret}
\author[2]{Giulio Costantini}
\author[2]{Stefano Zapperi}
\affil[1] {CNR-Consiglio Nazionale delle Ricerche, ISC, Via dei Taurini 19, 00185 Roma, Italy}
\affil[2] {Center for Complexity and Biosystems, Department of Physics,  University of Milan, via Celoria 16, 20133 Milano, Italy}

\begin{abstract}
Micro and nanoscale materials have remarkable mechanical properties, such as
enhanced strength and toughness, but usually display sample-to-sample fluctuations
and non-trivial size effects, a nuisance for engineering applications and an intriguing
problem for science. Our understanding of size-effects in small-scale materials has
progressed considerably in the past few years thanks to a growing number of
experimental measurements on carbon based nanomaterials, such as graphene 
carbon nanotubes, and on crystalline and amorphous micro/nanopillars and micro/nanowires.
At the same time, increased computational power allowed atomistic simulations to
reach experimentally relevant sample sizes. From the theoretical point
of view, the standard analysis and interpretation of experimental and computational data
relies on traditional extreme value theories developed decades ago for macroscopic
samples, with recent work extending some of the limiting assumptions of the original
theories. In this review, we discuss the recent experimental and numerical
literature on micro and nanoscale fracture size effects, illustrate existing theories
pointing out their advantages and limitations and finally provide a tutorial for analyzing
fracture data from micro and nanoscale samples. We discuss a broad spectrum of materials but
provide at the same time a unifying theoretical framework that should be helpful for 
materials scientists working on micro and nanoscale mechanics.
\end{abstract}
\begin{document}

\flushbottom
\maketitle

\thispagestyle{empty}

\section*{Introduction}

Understanding how materials fail is a multiscale problem of immense complexity that has fascinated and puzzled scientists and engineers for many years. Relevant processes range from the nanometer scales
where the atomic displacement and defect motion initiate irreversible deformation, to macroscopic scale where the deformation manifests itself in localized plastic instabilities and crack propagation. This intricate coupling between the scales reveals itself in the widespread observation of size  effects in materials strength, generally indicating that larger samples are easier to fracture. This general phenomenon was already noted by Leonardo da Vinci \cite{leonardo40}, who measured the carrying-capacity of  metal wires of varying length and observed that longer wires could sustain a smaller weight. This kind of fracture size effects is difficult to explain within a 
continuum mechanics framework: if the cross-section of the wire is constant, the stress is the same
regardless of the wire length. 

Fracture size effects can only be explained considering the role disorder, as originally
shown in the the pioneering work of Weibull in the 30's  \cite{weibull39}, which was 
based on extreme value theory (EVT)\cite{gumbel}: if a sample could be divided into a set of non-interacting subvolumes, the total strength would correspond to the strength  of the weakest volume, leading to the well-known Weibull distribution \cite{weibull39}. Additional 
support for this line of reasoning comes from fracture mechanics 
indicating that a material fails at a lower nominal applied stress when a flaw is present \cite{griffith20}.   
Coming back to Leonardo's wire, we can assume that longer wires are more likely to enclose longer
flaws and are thus bound to fail at smaller loads, on average. Unfortunately the issue is more complicated since the notion of independent subvolumes is often difficult to justify in practice:  
Flaws induce long-range stress fields in the material and therefore different sub-volumes may interact, invalidating the assumption of independence. As a result, it is not entirely clear if and why the Weibull distribution can be used to describe fracture statistics. For instance, compression
experiments in rocks indicate that the fracture strength does not vanish in the large size limit\cite{Weiss2014}, at odds with EVT but in agreement with depinning theories 
similar to those used to describe crystal plasticity \cite{zaiser2005,zaiser2006}.
Notice also that disorder is not always detrimental as implied by
EVT. In some cases, a strong disorder might even prevent crack propagation and ultimately
result in a stronger sample \cite{Tuzes2017}. Finally, size effects are commonly revealed experimentally but difficult to quantify with precision, since the strength probability distribution is dominated by its tails. Few scientists have the patience and the resources to repeat the same experiment on a large number of nominally identical samples. 

The current trend towards increasing miniaturization of samples and devices poses additional challenges to our understanding of  fracture size effects: Micro and nanomaterials often display rate and thermal dependent effects that could invalidate the weakest-link hypothesis which lies at the basis of EVT. Furthermore, concepts and ideas coming from continuum fracture 
mechanics, that are fundamental to understand macroscopic fracture, might not be
applicable to very small samples. Hence, while we can not rely on theoretical arguments 
that derive the Weibull distribution from the statistical properties of the flaws
present in an elastic medium \cite{Freudenthal1968,DLB,BealeDuxbury1988,Manzato2012,Bertalan2014},
the Weibull distribution is still commonly used
to fit and interpret experimental data, even at those small scales. 

Additional complications arise when we move from brittle to ductile materials. Plasticity in bulk materials has the common feature of being nearly size independent, so that the yield stress, the stress needed to initiate plastic deformation, is almost independent of the sample size. This statement held true until a few years ago, when materials scientists started to experiment with micron-sized samples. In their pioneering work, Dimiduk et al. performed compression experiments on single crystalline Ni micropillars, obtained by focused ion beam (FIB) machining \cite{uchic2004,dimiduk2005,uchic2009}. The results show a dramatic increase of  the yield stress as the pillar diameter decreases, but also strong sample-to-sample fluctuations and intermittent strain bursts during loading \cite{dimiduk2006}. Hence, micron-sized samples are not only stronger  than bulk ones, but also more erratic in their deformation. In single crystals, strain bursts were shown experimentally to follow a power law distribution, due to the collective dynamics of dislocations \cite{miguel2001,dimiduk2006,csikor2007}. These findings gave rise to a flurry of activity and similar size effects were recorded in a wide variety of crystalline \cite{greer2006,shan2008,greer2011} and amorphous materials \cite{wu2008,jang2011effects}.  Yet, a comprehensive and convincing theory accounting for size effects in micro and nanoscale ductile materials is still missing (see recent reviews \cite{uchic2009,kraft2010,greer2011,sethna2017}). 

This review is an attempt to respond to the pressing need of revisiting and extending existing size effects theories, suggesting guidelines for their application to experimental data. To this end, we first review recent experimental results revealing size effects in micro and 
nanoscale materials, ranging from carbon based materials such as carbon nanotube and graphene, to crystalline brittle and ductile metallic materials, and finally amorphous samples such as silica glasses and bulk metallic glasses. To help orient the reader through the vast experimental literature, we have compiled a list of experimental results in Table S1, including materials, sample size and statistical methods used. We then report on recent atomistic simulations results addressing the issue of size effects in those materials. To interpret all these data, we review 
classical EVT, discuss their limitations and propose some extensions appropriate to deal with rate and thermal effects which become increasingly relevant at the nanoscale. Equipped with this theoretical background, we propose general guidelines to interpret experimental data
and provide some concrete examples to illustrate our ideas.

\section*{Fracture at the micro and nanoscale: experiments}

\subsection*{Carbon based materials}
\begin{figure}
  \centering
\includegraphics[width=\linewidth]{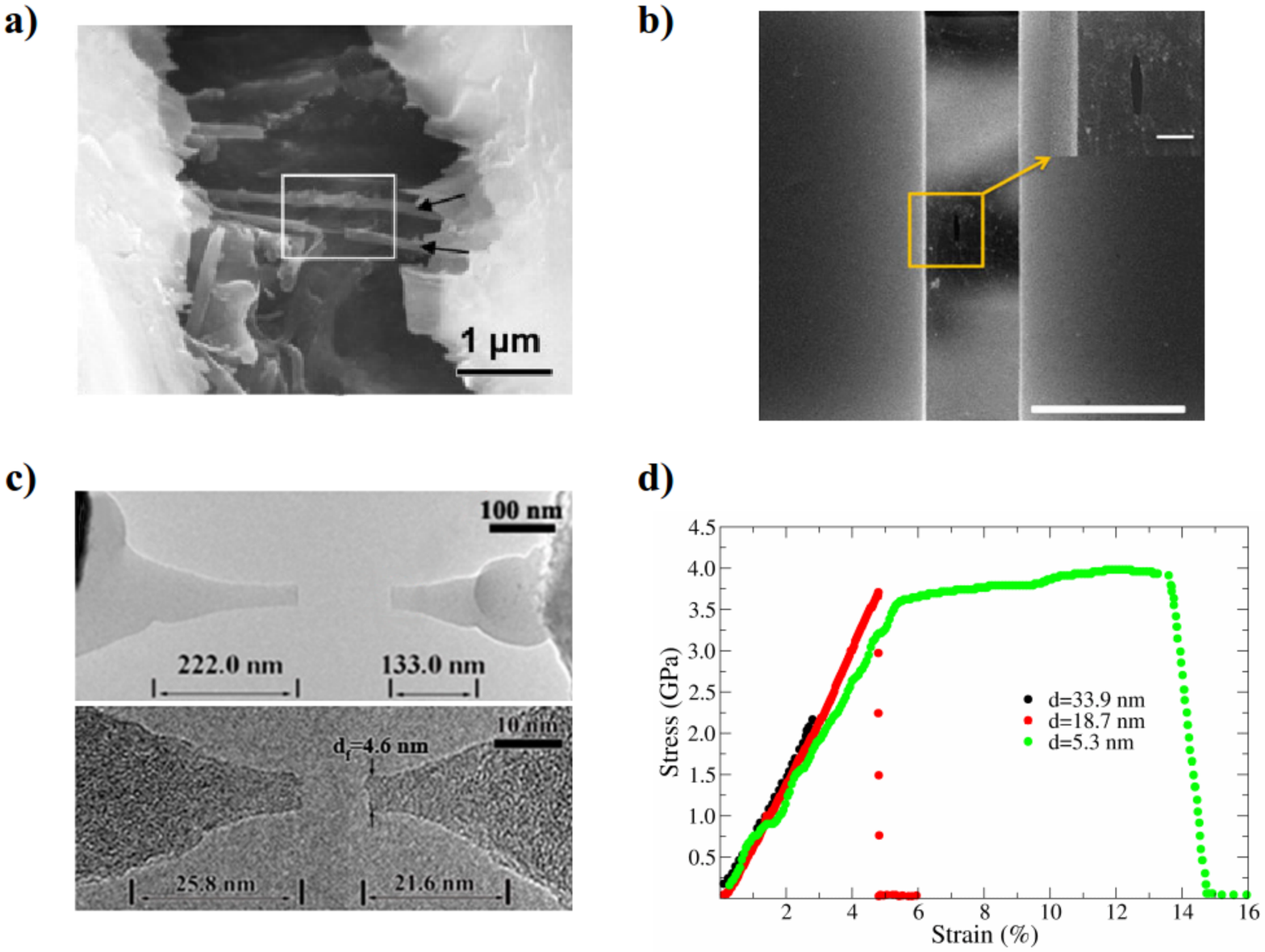}
\caption{{\bf Experimental failures.} Fracture behavior of nano materials with different structures and sizes.
{\bf a} Failure process of carbon nano-tubes during the in-situ tensile test. The CNT experience
different wall fracture as shown in the box \cite{chen2015load}.
{\bf b} SEM images of a graphene sample before a tensile testing. The pre-crack (in the box) was introduced by 
FIB cutting. The scale bar in b is 5 $nm$ while in its inset is 500 $nm$ \cite{zhang2014fracture}.
{\bf c} Morphology of SiO$_2$ nanofibers with different diameters after fracture under tensile loading. 
The upper and lower panels show nanowires of 33.9 $nm$ and 5.3 $nm$, respectively \cite{luo2015size}.
{\bf d} The stress-strain curves of three SiO$_2$ nanowires with different diameters \cite{luo2015size}. The smaller sample exhibits a ductile behavior.}
\label{fig:1}
\end{figure}

Due to their potential for technological applications, carbon based materials have played a major role in the past few years in the field of nanomaterials, with carbon nanotubes (CNTs) and graphene as the primary examples. These materials are interesting from the mechanical point
of view due to their extremely large strength and stiffness, as reviewed extensively \cite{mielke2007,jeon2016}. 
As an example of these experiments, we mention here early tensile tests on multiwalled CNTs using atomic force microscopy (AFM) revealing brittle fracture with extremely high tensile strengths in the range 20-60 GPa \cite{yu2000}.
This and subsequent AFM experiments on single filament multiwalled CNTs \cite{barber2005}
found size effects and failure stress distributions in reasonable agreement with Weibull statistics \cite{barber2005stochastic}. Weibull distribution has  been shown to successfully apply also to the fracture dynamics of CNT fibers \cite{zussman2005,deng2011,sun2012,zheng2010}, carbon fibers \cite{naito2012} and multiwalled WS$_2$ NT \cite{barber2005stochastic,kaplan2006}. Recent experiments on carbon fibers \cite{chae2015} and carbon fiber bundles \cite{hill2013} detected brittle failures, without exploring the statistical aspects. Among these results, we find of particular interest the work of Sun et al. \cite{sun2012} who recorded the distribution of tensile strength of CNT fibers as a function of strain rates and fiber diameters. While most of the tests on CNTs are based on AFM, other techniques involving force sensors and nanomanipulators have  been recently introduced \cite{jang2011}. It is also possible to produce CNT reinforced composites: Fig. \ref{fig:1}a shows experiments of the tensile deformation of multi-walled CNT embedded in a Al matrix \cite{chen2015load}. Experiments show that strain induces plastic deformation and microcrack formation in the composite.  In this setting, CNTs act as bridges and prevent crack growth increasing the fracture strength. Analyzing size effects and strength statistics in CNT composites is still an unexplored challenging task.

The other carbon based material that has recently excited much interest for its exceptional mechanical properties is graphene.  Testing its fracture properties is extremely challenging due to the difficulty in applying high tensile stresses in a controlled fashion on the sample \cite{Lee2008,ruiz2011,Kim2012,zhang2014fracture}. One of the few remarkable experiments considers the fracture of 
a bilayer graphene with notches of various lengths  (in the 66 nm – 2512  $\mu$m range) \cite{zhang2014fracture}.
The results, reported in Fig. \ref{fig:1}b, reveal a fast brittle fracture and a reduced strength with respect to 
the intrinsic value. This can be explained by the classic Griffith's theory  that provides an expression for the 
crack length dependence of the failure strength (see the
theoretical section of this review). It would be interesting to understand how a collection of 
randomly placed nanoscale cracks or vacancies would affect the strength, but this issue has still to be investigated experimentally. 

\subsection*{Silicon based  materials}
Among single crystals, silicon holds a preminent role, as it is one of the most common materials in electronic devices and micro and nanoelectromechanical systems (MEMS/NEMS).  Si and Si-related single crystals
are brittle at room temperature and statistical analysis based on the Weibull law is commonly used to describe the  observed scatter in bending or tensile strength values\cite{bhushan2017}. Already in 1985 Petrovic and coworkers  \cite{petrovic1985} reported tensile experiments on silicon carbide (SiC) whiskers whose diameters were in the range of 4 to 6$\mu$m, and lengths  approximately 10mm. Measured strength values exhibited a significant
range with a statistics in good agreement with the Weibull theory. More recently,  SiC dog-bone shaped specimens were  subject to microtensile uniaxial experiments, demonstrating the ability of the Weibull distribution to be used to predict the strength response\cite{nemeth2007}.  Later, tensile test devices were designed and developed for operation not only in ambient air but also in a FIB and a SEM \cite{fujii2013}. The width and the thickness of FIB-fabricated nanowires were varied within the ranges from 57 and 235 nm to 221 and 444 nm respectively, while those of annealed nanowires ranged from 149 and 314 nm to 263 and 418 nm. Results indicated brittle failure during elastic deformation  in the entire set of wires tested. Bending tests on micro and nanofilms corroborate the brittle nature of Si and related composites. Nanometer-scale Si double anchored beams, with widths from 200 to 800 nm and a thickness of 255 nm, were loaded by means of AFM \cite{namazu2000} or  a depth-sensing nanoindenter with a harmonic force \cite{li2003}. In the first case, the bending strength distribution followed the Weibull statistics, with clear size effects. The same qualitative behavior was reported\cite{sundararajan2002} for Si and SiO$_2$ nanobeams with a 6$\mu$m length and widths ranging from 200 to 600 nm deformed in bending using an AFM. Single crystal 3$C$-SiC together with ultranano-crystalline diamond (UNCD) and amorph one carbon (ta-C) micro-specimens were tested in membrane deflection experiments \cite{espinosa2006},  confirming the
validity of Weibull theory in predicting the specimen strength when 
the volume was changed by about two orders of
magnitude. At higher temperature, Si becomes plastic and the brittle-to-ductile transition temperature decreases with sample size \cite{kang2017}, as shown  for single crystal silicon (SCS) beams with widths of 720nm to 8.7$\mu$m in  thermo-mechanical bending tests \cite{kang2013}, and for microbeams in tensile regime \cite{uesugi2015}.
 However, the matter of the brittle or ductile nature of Si NWs at room temperature is still rather controversial \cite{hoffmann2006,gordon2009,zhu2009,zhang2010,steighner2011,kim2011,kizuka2005,han2007,zheng2009,wang2011,ostlund2009}. In situ TEM experiments performed under both uniaxial and bending conditions for ultrathin Si NWs down to $\sim$ 9 nm diameter, revealed an extremely rich behaviour subtending the NW fracture mechanics, simultaneously influenced by the NW dimensions, loading conditions and stress states
 \cite{tang2012mechanical}.

 Generally speaking, the capability of Weibull statistics to predict localized strengths at micro and nanoscales as it does for  bulk ceramics has also been accurately reported in \cite{jadaan2003}, considering the four most widely used materials  in MEMS/NEMS fabrication:   SiC, SCS, silicon nitride and polycrystalline silicon (poly-Si). Poly-Si thin  films,  indeed, constitute a key structural component for various microelectronics devices\cite{french2002} indeed. At room temperature, poly-Si thin films fail in a brittle  manner by cleavage \cite{boyce2010} and Weibull parameters may describe sufficiently well the probability of failure of a variety
of  geometries with different stress distributions, provided that the active flaw population is properly identified \cite{boyce2007}. This was shown in uniaxial tension tests of poly-Si dog-bone specimens with uniform cross-section, with a central hole, and with symmetric double notches 3.5$\mu$m thick, with widths of either 20 or 50 $\mu$m \cite{bagdahn2003}. The same has been observed in the case of micro specimens with various geometries of circular/elliptical perforations centrally located at the gauge section \cite{mccarty2007}. The fracture strengths of 40 and 240nm poly-Si thick films have been demonstrated to decrease with the film thickness \cite{vayrette2016}.  These  size effects were shown to be related  to the polycrystalline
nature of the material and to microstructure difference controlled by the fabrication process \cite{chasiotis2003}.

Size effects in the fracture of nanoscale and microscale amorphous materials follow remarkably similar trends as those observed in crystalline materials, and are routinely described by Weibull statistics \cite{andersons2002}. A clear example is provided by silica glass (SiO$_2$) samples that   can be manufactured into nanowires \cite{brambilla2009ultimate} or even dog-bone shaped nanofibers with diameters going down to 1nm \cite{luo2015size} (see Fig. \ref{fig:1}c).  Tensile deformation tests of these samples display a brittle-to-ductile transition at room temperature when the sample diameter goes below
18nm \cite{brambilla2009ultimate}. Reducing the sample size leads to a dramatic increase in 
tensile strength  and accumulated plastic strain with respect to bulk samples. 
 This is illustrated in Fig. \ref{fig:1}d comparing the stress-strain
curves for samples of different diameters. 
The current interpretation of this experimental observation is that the effect is due to the increasing role played by enhanced atomic diffusion at the boundary of the sample \cite{luo2015size}.

\subsection*{Metallic materials}
Statistical fracture size effects in macroscopic samples have  been traditionally investigated in 
brittle and quasi-brittle materials, whereas size influence was barely considered 
in ductile metallic materials \cite{sevillano2001,greer2011,uchic2009,kraft2010}. 
Materials, however, can be brittle or ductile depending on size, shape, specimen preparation, experimental conditions such a strain rate or temperature,  and even on the type of the experiment performed. A paradigmatic example is copper, which is the  most widely studied among ultra-fine grained and nano-crystalline materials \cite{khan2008}.  As for many other single-crystals, Cu exhibits  plastic behavior whose onset is intimately tied to presence and depinning of dislocations. Plasticity in Cu specimens 
at the micro and nanoscale has been detected in a wealth of different loading conditions and sizes. Uniaxial tensile tests on single-crystal  Cu microneedles \cite{kiener2008}, thin films \cite{lin2010} and nanopillars with diameters between 75 nm and 165 nm \cite{jennings2010microstructure}, have been performed in \emph{in-situ}  apparatuses, i.e. with the help of  a scanning (transmission) 
electron microscopy to achieve the high-resolution detection of  deformational fields \cite{haque2002}. Furthermore, a surprising crystalline-liquid-rubber-like  behavior in Cu nanowires was reported in ref.\cite{yue2013}, with a retractable strain of the fractured wires that can approach over 35$\%$. Plastic behavior has been assessed through bending experiments as well; these include  microsized cantilevered 
beams \cite{motz2005,kiener2006} and thin metal films on silicon substrates \cite{florando2005}, deflected using a nanoindenter, and bulge experiments \cite{vlassak1992} 
down to the nanoscale domain \cite{wei2007, merle2012}. 

From the late fifties of the last century, copper whiskers were known exhibit brittle behavior.
Under tensile loading, whiskers of few millimeters in length and few microns in diameter ``suddenly snapped, without any observable amount of plastic deformation'' \cite{brenner1956,brenner1958}. Fracture size effects were present in almost all samples, as the strength increased when the length or the section decreased.  The same brittle behavior (no plastic necking) has been observed in nano-whiskers with diameters  between 75 nm and 300 nm by attaching a micromanipulator to one end of the wire and pulling on it \cite{richter2009}. In general, this size-induced ductile-whisker transition appears to be common  in  many crystals (Ag, Fe, Au, Pd, Ge, Si, Zn and Cd) \cite{greer2011,dimiduk2005}, 
with the two phases well characterized  at the atomic level (whiskers can be considered dislocation-free) and by distinct size-dependent behaviors. 

Plastic size  effects manifest themselves quite generically at micron and sub-micron scale where the yield strength and flow stress increases by reducing the specimen size  \cite{kraft2010,greer2011,sevillano2001,hemker2007}. Although their specific plasticity mechanisms is 
still debated, the generic mechanism is believed to be  dislocation-mediated  \cite{gianola2006,gupta2017,greer2011,chen2017,merle2012}. These plastic size effects 
are associated to intermittent discrete slip events separated by elastic loading segments in the stress-strain curve, as shown by uniaxial compression experiments performed on nanopillars and micropillars made out of Cu \cite{jennings2011tensile,jennings2011},
Ni \cite{dimiduk2005,shan2008,uchic2004}, Au \cite{greer2006}, W \cite{wang2015}, Mg \cite{lilleodden2010}, Al \cite{kunz2011}, Mo, Ta and Nb \cite{kim2010}, and in dislocation
dynamics simulations \cite{csikor2007}. Despite the large scatter in  yield strength values, however,  the observed  stress values do not approach the magnitudes reported for brittle fracture in whiskers, although the diameters of the samples significantly overlap.

To make the playground even richer, a remarkable ductile-to-brittle transition has been observed in uniaxial tensile tests of Au nanowires  (diameter between 8 and 20nm)  containing angstrom scale twins \cite{wang2013}. Moreover, recent experiments performed on freestanding Cu thin films, in micromachined tensile frame at  elevated temperatures, revealed that failures of smaller samples (385 nm) occur at the onset of the elastic-plastic transition, pointing out that  the ductile-to-brittle temperature may reduce with the sample size  \cite{sim2013}. Using a micromachined silicon tensile frame with integrated heaters, Sim \emph{et al.} reported a large decrease in yield strength at elevated temperature for freestanding Au thin films of width varying from 450 to 960nm \cite{sim2014}. 
They also observed an inverse size effect where the yield strength at elevated temperature decreases with decreasing temperature.  Thermally-activated deformation  mechanisms at the micro/nanoscale
are currently being addressed, and systematic investigations of the effect of size on mechanical strengths as a function of temperature have now been conducted for a number of
 materials \cite{kang2017}. Moreover, compression of Cu nanopillars with diameters ranging from 75 up to 500 nm \cite{jennings2011} and  uniaxial tensile 
test of 200 nm-thick film  \cite{vayrette2015}, have shown that  the apparent fracture strain during the ductile phase  is significantly affected by the  strain rate \cite{gorham1991}.

Another class of metallic materials that has been actively investigated is represented by bulk metallic glasses (BMG). Under uniaxial tensile load, most monolithic BMGs fail catastrophically without any plastic deformation at  room temperature \cite{sun2015fracture}. Plasticity can be observed only in a few specific cases, such as in dynamic testing with high strain rates or in nanoscale size samples. Under stable loading geometries, such as uniaxial compression, however, BMGs often display  plasticity before final fracture.  The strength and the shear strength of BMG are usually about  much smaller than the theoretical strength \cite{wang2012elastic}. The deviation from the theoretical value is attributed to the existence of manufacture flaws or other structural defects.
Structural investigations reveal that BMG fail by accumulation and localization
of repeated plastic shear events, softening the material up to the formation of 
a shear band inducing catastrophic failure \cite{greer2013}. This scenario can be framed into the weakest link paradigm if we associate the failure with the appearance of the first shear band overcoming 
a critical size. 

In general, compression experiments on BMG micropillars report marked increases in yield stress and strength  over the corresponding bulk values. Examples of this behavior include Zr-based metallic glass samples with diameter going down to 100nm \cite{jang2011effects}, where the yield strength increases as the diameter decreases down to 800 nm, and then remains at its maximum value of 2.6 GPa. Earlier studies on other Zr-based metallic glasses with diameters on the micron range display  localized plastic flow, strain bursts during the deformation curve and a large increase (25-86\%) of the yield stress with respect to the bulk values \cite{lai2008bulk}. 
Other measurements are performed in Pd-based metallic glasses with nominal diameters of 2-20 $\mu$m using quasi-static room temperature compression, showing a 9\% increase in yield stress with respect to the bulk values \cite{schuster2007bulk} and in Mg-based micro scale metallic glasses where the observed yield stress increment goes up to (60\%-100\%). More recent measurement on Zr-Cu-Al-Ni metallic glasses reported also a strain rate dependence of the yield stress \cite{chen2015microstructural}. In these and other studies \cite{calvo1989,ocelik1987,zhao2008,wu2008,han2009,shamimi2008,yao2008,lee2010}, the Weibull distribution is used to quantify fluctuations and size effects, confirming the relevance of the weakest-link scenario for the failure of BMG  (for a review see also Ref. \cite{greer2013}.)

\section*{Fracture at the micro and nanoscale: atomistic simulations}

Molecular dynamics (MD) simulations are a tool complementing experiments by allowing to investigate and elucidate the microscopic properties of material failure, often revealing details hidden to bulk measurements. Thanks to the improvements in computational power in the last decades, MD simulations have been used as a predictive tool for the macroscopic response,  improving 
experimental design. The main limitation of MD simulations in general, and in their application to fracture in particular, lies in the accessible length and time scales. In practice, it is only possible to simulate a relatively small sample for a limited time interval and therefore experimentally relevant deformation rates are out of reach for MD simulations. Furthermore, MD simulations rely on empirical interatomic potential that are typically validated using quantum
mechanical methods or experiments. Most of the potentials are often adapted to the small deformation
regime and could therefore provide spurious results in simulating fracture unless special care is taken \cite{pastewka2008}. 

CNTs are among the most simulated materials because of many well-established interatomic carbon potentials \cite{brenner1990,van2001}. Earlier simulations considered single-walled carbon nanotubes (SWCNTs),  6.4 $\AA$ in diameter and   
100 $\AA$ in length,  with the goal of understanding the influence of topological defects on the mechanical properties\cite{yang2007size}, in agreement with earlier works \cite{lu2005}. 
Results showed large fluctuations in the fracture strength due to the randomly distributed defects on the single-walled nanotubes (SWNTs) surface. The statistics
was found in good agreement with the Weibull law, as in earlier studies\cite{bhattacharya2006}. 
Other MD simulations demonstrated that the failure strain of CNTs  subject  to tensile deformation rises by increasing the strain rate and decreases with increasing temperature \cite{yakobson1997}. Wen \emph{et al.}  proposed a thermally activated model connecting the  failure strain to strain rate, temperature and CNT length, comparing the predictions with MD simulations \cite{wen2009}. Simulations with a wider temperature
range, from 300K to 2700K, showed that at high temperatures damage accumulates before failure, while at low temperature brittle fracture always occurs shortly after the failure of the first bond \cite{brenner2002}.  Along the same lines, more recent extensive numerical simulations of relatively large SWCNTs (diameters ranging from 0.63 nm to 4.7 nm,  and  length equal to 10.6 nm) at varying temperatures (300 to 2400 K) 
showed tensile ductility for large CNT diameters, over a wide temperature range (500–2400 K) \cite{tang2009molecular}. 
For smaller diameters, SWCNTs display brittle fracture due to the  strong localization of incipient defects, even  at high temperatures.
The linear increase of the elastic limit of SWCNTs with decreasing temperature is microscopically rooted in defect nucleation and  
dynamics, promoted by the interplay of thermal and strain energies.

\begin{figure}
  \centering
\includegraphics[width=\linewidth]{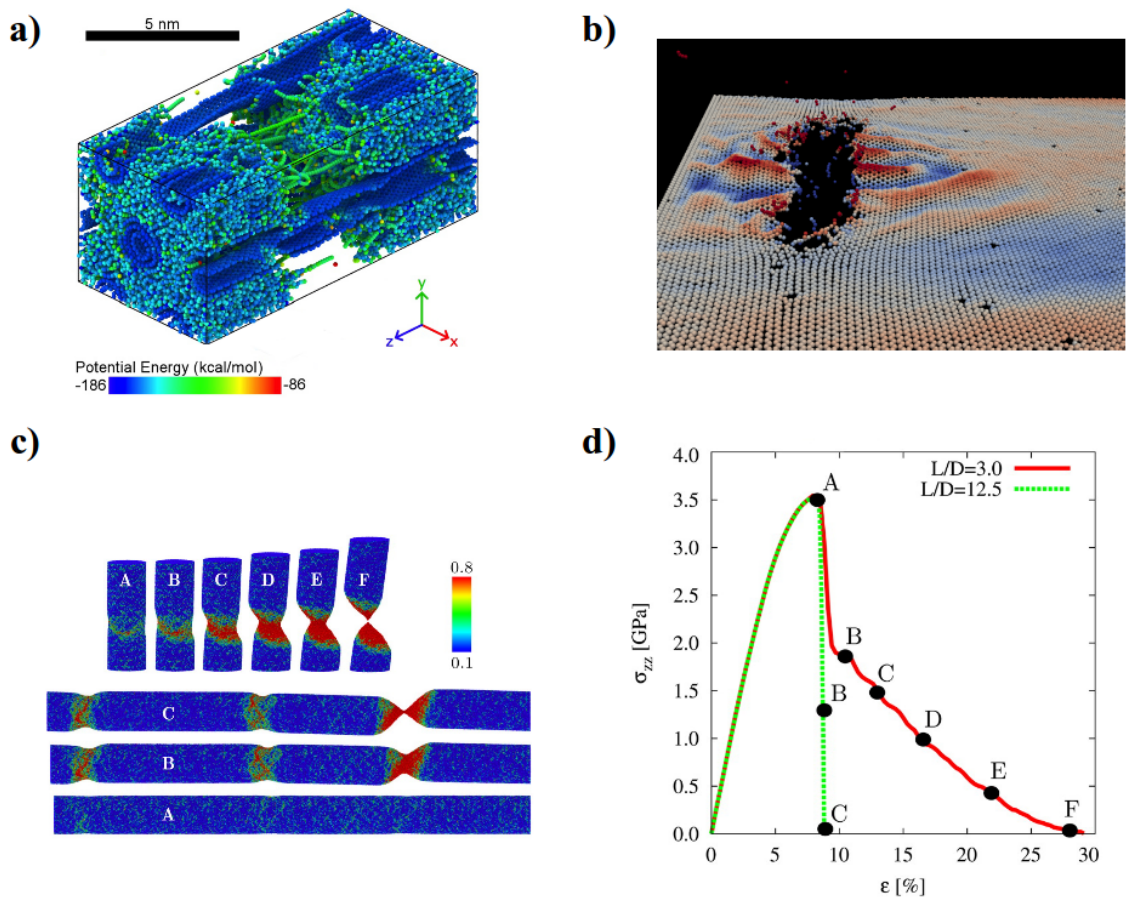}
\caption{{\bf MD simulations.} Images of different structures undergoing fracture from Molecular Dynamics Simulations. {\bf a} Multi-wall nanotube array with nominal crosslinking fraction of 20\% strained 0.50 in the axial
direction \cite{jensen2016simulation}.
{\bf b} Failure of a graphene sheet composed of N$=50\cdot 10^3$ atoms, with a vacancy concentration of 0.1\% \cite{sellerio2015}.
{\bf c} Snapshots from simulations showing the local atomic shear strain 
of two Cu$_{64}$Zr$_{36}$ glass nanowires subject to an uniaxial tensile loading along the major axis \cite{sopu2016brittle}.  The panels A to F (upper sequence) and A to C (lower sequence) have an aspect ratio of 3 and 12.5 respectively. The color scale indicates the atomic local shear strain. {\bf d} The stress-strain curves of the samples in the panel c. The points A-F and A-C on the curves show the corresponding deformation stages of the samples.}
\label{fig:2}
\end{figure}

Among their valuable properties, the extremely high stiffness and strength make the CNT the best candidates as reinforcement nanodevices in 
next-generation structural composite materials. To this end, it is crucial to optimize the load transfer between individual CNTs, either directly or through the matrix (see Figs. \ref{fig:1}(a) and \ref{fig:2}(a)). Jensen et al. considered an 
amorphous carbon (AC)   matrix  and simulated CNT/AC composites\cite{jensen2016simulation}. CNT/AC composites  were arranged in three  different configurations: SWCNTs in a uniformly spaced array, multi-wall nanotubes (MWCNTs) in a uniformly spaced array, and SWCNTs in an array of bundles.  Chemical crosslinking was induced by increasing the amount of bonds between the CNTs and the matrix, allowing to probe the trade-off between weakening the CNTs and improving the load transfer. 

Large-scale MD simulations of tensile deformation and failure of graphene sheets for a wide range of sample sizes,
vacancy concentrations, temperatures and strain rates, have led to a new theory for thermally activated rate-dependent fractures in brittle materials \cite{sellerio2015} (see Fig.\ref{fig:2}(b)).  The theory was inspired by single-molecule pulling models \cite{dudko2006} and generalized the classical extreme values statistics to thermal and rate effects.
Theory and simulations showed that the failure strength decreases with raising temperatures and sample size, while it increases with  strain rate.  The observed deviations from the weakest link hypothesis could all be resolved within this framework,  highlighting its predictive power for generic brittle materials.

Si and Si-based materials have been the subject of extensive MD simulations in the last decade. One important aspect of such numerical approaches is the volume scaling effect, and its interplay with the temperature and strain rate dependence, which has strongly motivated the development of new testing methods to study the coupling effects. MD simulations of tensile tests of Si nanowires (NWs) with diameters between 2 nm and 7 nm were reported in a temperature range from 100 K to 1200 \cite{kang2010}. Stress-strain curves highlighted important thermal effect on fracture strength, decreasing monotonically as temperature increased. At the same time, the failure stresses were found to increase with strain rate. The microscopic fracture mechanism was shown 
to depend not only on temperature but also on the NW diameter, with thinner NWs exhibiting a shear failure mechanism  and  ductile behavior. On the other hand, larger NWs were observed to fracture by a cleavage mechanism, originating at the surface, leading to brittle failure. The onset of intrinsic plastic size-effects was shown to be in qualitative agreement with experiments, although the critical diameter was substantially smaller than that in tensile experiments. Similar results were reported for ultrathin Si NWs, studied by uniaxial tension and bending \cite{tang2012mechanical}. Tensile tests revealed brittle fracture due to the nucleation and propagation of a single crack and cleavage along the (111) planes. Decreasing the NW diameter, the tensile strength tended to increase from 4.4 to 11.3 GPa. Under bending, the Si NWs showed considerable plasticity and for low strains no crack could be detected so that the Si NWs could be repeatedly bent. At higher bending strains, a crack, produced on the tensed side, propagated quickly, allowing the bending strain to  localize
around the crack tip. Yet, enormous plastic deformation was observed on the compressed side, due to the formation of amorphous structures.

Material properties at the nanoscale cannot be directly extrapolated from their bulk counterparts, and are often hard to achieve using the available material testing techniques.  Therefore, in many cases MD simulations provide the appropriate tool to reveale unexpected underlying small-scale-specific mechanisms.
Amorphous silica NWs with diameters from 2.23 nm to 10.23 nm and  a gauge length of 12.6 nm were simulated  in uniaxial tension tests \cite{zhang2015size}. Stress–strain curves showed that Young modulus, peak stress and ultimate strength all increased as
the sample diameter was reduced, but remained always smaller than values for bulk silica. The failure mode, however, showed  
very different behavior going from small to large diameters: NWs with the smallest diameter ($D=2.23$ nm) fractured  
only by localized necking and void growth did not affect the overall failure. For larger diameters ($D\geq 2.23$nm), 
growth and coalescence of voids occurred inside the NWs, and cracks nucleated and propagated on the
surface of NWs.

In the case of metallic glasses, MD simulations were performed for Cu$_{50}$Zr$_{50}$ NWs  
\cite{zhou2015size}. Compression and tension tests were used on cylindrical specimens with diameters $D$ ranging from 8 to 45 nm and a ratio of height to diameter of 3.  It was found that shear localization ruled  deformation in the case of sub-micro samples, yielding a $D^{-1/2}$ scaling law, whereas homogeneous deformation mode led to a $D^{-1}$ dependence of the yield strength for nanoscale samples. On the basis of a  theoretical model involving surface stress and Mohr–Coulomb criterion, the authors were able to estimate the critical length scale corresponding to a transition from shear localization to homogeneous deformation \cite{zhou2015size}.
More recently, two Cu$_{64}$Zr$_{36}$ NWs with a diameter of 20 nm and aspect ratios of 3 and 12.5, respectively, were simulated under 
uniaxial loading at 50 K and a constant strain rate of $4\cdot 10^7 s^{-1}$  (see Fig.\ref{fig:2}(c)-(d)) \cite{sopu2016brittle}. 
These simulations could provide an atomistic understanding of the deformation mechanism, separating the size effects from 
the contribution to plasticity due to the aspect ratio.  The two NWs did show substantial different plastic regimes once the maximum stress was 
reached. Although they both exhibited very similar elastic and plastic deformations localized in shear bands, only the high aspect ratio NW failed catastrophically owing to the initiation and propagation of one shear band.   A theoretical model was developed, providing a qualitative interpretation of the brittle-to-ductile transition.
Finally, metallic glasses with structural flaws were simulated by  considering
notched and unnotched Fe$_{75}$P$_{25}$ nanocylinders\cite{gu2014mechanisms}. The authors studied the deformation mechanism and failure modes under uniaxial tension, demonstrating that the brittle propagation of cracks in the notched samples was due to  
void nucleation, growth, and coalescence. In unnotched samples, shear band formation led to failure in a direction not orthogonal to the loading one. In general, structural flaws considerably reduced the sample failure strength and affected both crack initiation and rupture mode, even in the absence of discrete microstructural features.

\section*{Statistical theories of fracture}

Extreme value theory (EVT) deals with the statistical properties of the extremes (i.e. the maximum or the minimum) of $N$ identical independent random variables. The central result is that in the large $N$ limit, the distribution of
extremes has a limiting form falling into three general classes: the Weibull, the Fr\'echet or the Gumbel distribution \cite{fisher1928,frechet1928,gnedenko1943}. The Weibull and Fr\'echet distributions are intimately related, but the Weibull distribution is mostly used in cases that deal with the minimum rather than the maximum, hence the distributions of failure stresses 
in materials are believed to be either of the Weibull or of the Gumbel type (see Box 1). 

The connection between EVT and fracture is provided by the ``weakest link hypothesis'': assuming that a sample of volume $V$ can be subdivided into $N$ representative elements of volume $V_0$, the fracture strength is determined by the 
smallest failure stresses of the elements, resulting in an extreme value distribution. This can be seen considering that the probability at a set of $N=V/V_0$ elements with random failure stresses survives at a stress $\sigma$ is equivalent to the probability
that all the elements survive, yielding $\Sigma_V(\sigma) = \Sigma_0(\sigma)^{V/V_0}$, where $\Sigma_0(\sigma)$ is the survival distribution of the single element. Since by definition 
$\Sigma_0$ is a decreasing function that is always less than one for $\sigma>0$, when $N$ increases the average failure stress must decrease.

The straightforward application to EVT to fracture rests on several assumptions that is sometimes hard to justify, leading to an intense debate~\cite{danzer2007,dieter2002,Doremus1983,Jayatilaka1983,Manzato2012,nohut2012,bazant1991,rozenblat2011,basu2009}. 
In particular, EVT invariably assumes that the material volume can be decomposed in a set of statistically independent elements or that it has a population of non-interacting crack-like defects, so that global failure occurs as soon as the weakest of these defects starts to grow.
This assumption works well for some brittle materials such as glasses or ceramics where the distribution of fracture strength can be derived from the distribution of flaw sizes \cite{Freudenthal1968,DLB,BealeDuxbury1988,Manzato2012,Bertalan2014}. To demonstrate this,
we should consider the stability of a crack in a linear elastic material, as first derived by Griffith in his pioneering work \cite{griffith20}. According to Griffith's theory, a crack of length $w$ subject to a normal stress $\sigma$ is stable as long as  
\begin{equation}
K = \sigma Y w^{1/2} \leq K_{Ic},
\label{eq:StressIntensity}
\end{equation}
where $Y$ is the geometry factor of the crack, and $K_{Ic}$ is the critical stress intensity factor of the material.  If we known the distribution $P(w)=e^{-h(w)}$ that a volume element $V_0$ does not contain any crack longer than $w$, then we can invert Eq. \ref{eq:StressIntensity} to derive the survival distribution as
\begin{equation}
\Sigma_V(\sigma) = \exp\left(-\frac{V}{V_0} h(K_{Ic}^2/\sigma^2Y^2) \right).
\label{eq:DefectDensity}
\end{equation}
If the crack length distribution is a power law with exponent $\gamma$, then 
$h(w) \sim w^{-\gamma}$, which leads to a Weibull distribution of fracture strength with modulus $k = 2\gamma$ \cite{Freudenthal1968,Bertalan2014}. When the crack length
distribution is exponential, the survival distribution asymptotically becomes of the
Gumbel form \cite{DLB,BealeDuxbury1988,Manzato2012}. 

Quasi-brittle materials such as paper~\cite{Alava2002}, granite~\cite{lockner1991quasi,garcimartin1997}, 
bone~\cite{Zioupos1994203}, wood~\cite{garcimartin1997,Tschegg2000}, and composites~\cite{yukalov2004,sornette1995}, typically accumulate damage before fracture. Hence, the weakest defect would not necessarily dominate the fracture properties and multiple defects would also interact via long-range elastic fields so that the assumption of statistical independence becomes questionable. 
Numerical simulations of network models for fracture \cite{Alava2006} indicate that in the large size limit the survival distribution can be successfully rescaled so that, even if volume elements
are interacting, the failure distribution converges to the Weibull \cite{Bertalan2014} or Gumbel class \cite{DLB,BealeDuxbury1988,Manzato2012} depending on the type of disorder. The asymptotic distribution, however, does not simply reflect the initial disordered configuration of the sample but involves a correction due to accumulated damage \cite{Manzato2012,Bertalan2014}.

In experiments, testing for the independence of individual sub-volume elements can be performed 
empirically by data collapse, as illustrated in Fig. \ref{fig:3}. In particular,
for independent sub-volumes $\ln \Sigma_V /V$ should not depend on $V$ so that rescaled survival curves obtained from samples with different volumes should all collapse into a single master curve. Successful rescaling also implies that the average failure stress depends on the volume, as shown in Fig. \ref{fig:3}c. If we consider a survival distribution obtained from individual volume elements whose failure stresses are not statistically independent, we can not rescale and collapse the data, as shown in Fig. \ref{fig:3}b. Hence, volume rescaling of the survival distribution is a crucial test for the applicability of EVT to fracture data. 

A straightforward application of EVT to micro and nanoscale materials is even more problematic, since we can not rely on the large scale asymptotic limit. Despite this fact, fitting size effects and strength distributions with  Weibull-type laws is extremely common in micro and nanoscale materials mechanics. The main issue in small scale samples
is that the volume $V_0$ of representative element is expected to be small and therefore sensitive to thermal fluctuations. 
This implies that a meaningful statistical theory for fracture size effects should include the effect of temperature and strain rate. Intuitively, if a representative element can fail due to thermal activation then rate effects also become important since the slower the rate, the likelier it is that a critical thermal fluctuation will lead to failure.

A generalization of EVT taking into account strain rate and temperature is discussed in Box 1 \cite{sellerio2015}. The theory replaces the search for the weakest link by the calculation of the first link to break. In this process, extreme value statistics is
changed into a first-passage problem. The interesting observation is that the general structure of the survival probability remains the same in the EVT case, namely $\Sigma(\sigma)=\exp[- V/V_0 f(\sigma,T, \dot{\varepsilon})]$, where $f(x,y,z)$ is a suitable
function of stress $\sigma$, temperature $T$ and strain rate $\dot{\varepsilon}$. This could explain why one can often fit experimental data for micro and nanoscale samples with EVT (see Fig. \ref{fig:3}d). Provided that temperature and rate are held constant, the expected size effect law will resemble the predictions of EVT (see Fig. \ref{fig:3}d). Yet, for a more precise determination of size effects it is imperative to analyze the role of strain rate and temperature as it has been shown in the case of graphene simulations \cite{sellerio2015} (see Fig. \ref{fig:3}e
and \ref{fig:3}f). The theory suggests that the relevance of rate and thermal effects can be simply estimated computing 
a dimensionless parameter $\Lambda = \frac{\omega_0}{\dot{\varepsilon}}\sqrt{EV_0/(k_BT)}$, where $E$ is the Young modulus, $k_B$ is the Boltzmann constant, $V_0$ is the volume of the representative element and $\omega_0$ is its characteristic frequency. Only when $\Lambda \ll 1$, thermal and rate effects can safely be ignored.

\section*{A data analysis toolbox for fracture statistics}

When confronted with a set of data for the fracture strength of different samples, the key question to address is which statistical distribution best represent the data and how does this distribution  change with the sample volume. In ideal conditions, we would need a large set of nominally identical samples for each different volume, but unfortunately this is rarely available. In most cases, we have access to a relatively small set of samples, possibly of varying volumes. The question is then to decide the best strategy to analyze them, avoiding possible pitfalls. In Box 2, we provide a set of technical guidelines on how to perform this analysis in practice depending on various scenarios of increasing complexity. In the first scenario (case i in Box 2), we consider a set of samples of equal volume $V$ and assume that we can safely disregard rate and thermal effects (i.e. $\Lambda \ll 1$). Under these conditions, the best strategy is to perform a simple fit of the survival distribution function obtained from the data   $\Sigma_V^{exp}(\{\sigma_i\})$ using a suitable function obtained from EVT. 

In practice, $\Sigma_V^{exp}(\{\sigma_i\})$ can be constructed by rank ordering the experimentally obtained failure stresses: $\{\sigma_i\}$, where $i=1, ... N$ is the rank order, from smaller to larger, and then plotting ($\sigma_i$, $1-(i-1/2)/N$). This procedure provides an estimate of the survival distribution that is free from any  binning. Notice, that we have to be extremely careful in applying the same  procedure to the common case in which the experimentally tested samples do not share the same volume $V$. If we construct a survival distribution in such a case, we would be  mixing together the distributions related to different volumes $V_i$. Hence, we can {\it not} use the EVT prediction derived for identical volume samples to fit the distribution, although this was often incorrectly reported in the past literature. The only viable option to compare data obtained from samples with different volumes is to use the maximum likelihood method (see case ii in Box 2).

Two other scenarios occur when the rate and thermal effects can not be neglected in the fracture
process. In those cases, one should again distinguish between the case in which the volume is
the same for all samples (case iii in Box 2) or different for each sample (case iv). As in the
previous scenarios, equal volume samples can be analyzed either by least square fitting  or by 
the maximum likelihood method, while samples with different volumes can only be analyzed with 
the second method. In both cases, the reference theory is not classical EVT but a temperature
and rate dependent version of it, like the one reported in Box 1.

\begin{figure}
  \centering
\includegraphics[width=12cm]{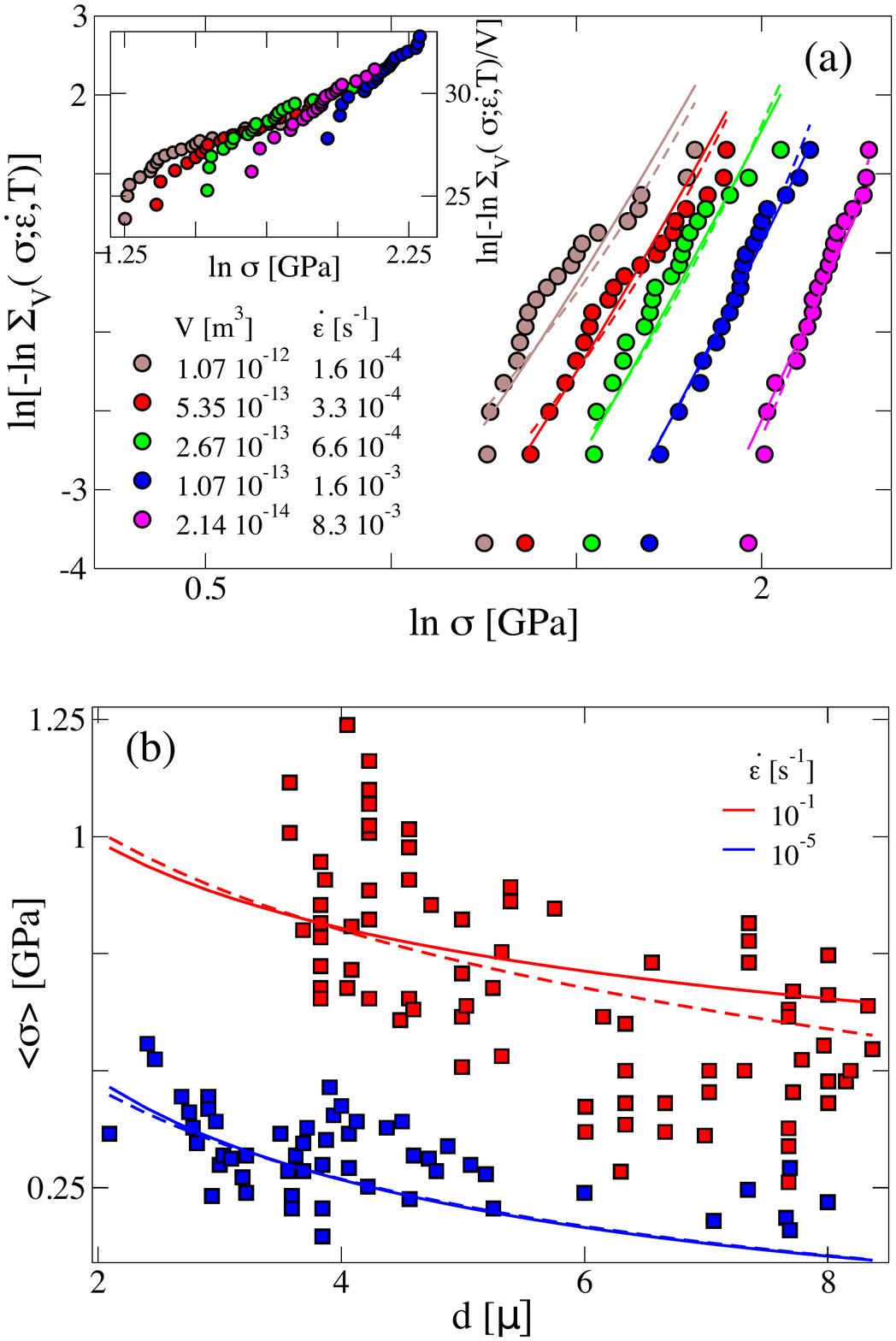}
\caption{{\bf Examples of EVT parameter estimations from experimental data}. {\bf a} Main panel: survival distribution function obtained from the experimental failure stresses of ultrahigh strength PAN-based carbon fibers (IM600) (filled symbols) \cite{naito2012}. Experiments are conducted at different volumes and strain rates. Since we have data from many samples for each value of $V$ and $\dot{\epsilon}$, the least-square method can used to fit simultaneously the survival distribution functions using the rate dependent theory under the Weibull (solid lines) or Gumbel (dashed lines) hypothesis. Although experiments are performed at different strain rates for each volume $V$, $\Lambda \ll 1$ and the survival curves collapse when rescaled by the corresponding volumes. {\bf b} Failure strengths obtained from tensile tests performed on multiwalled carbon nanotube fibers (filled squares) \cite{sun2012}. Experiments are conducted at two different strain rates and for various diameters $d$. Since the data are obtained at different volumes, the maximum likelihood method has been used to determine the distribution parameters.  Solid lines represent the Weibull average failure strength calculated according to the estimated parameters, whereas dashed lines correspond to the Gumbel case. Here for high strain rate, $\Lambda >1$ so rate effects can not be neglected.
}
\label{fig:4}
\end{figure}

In Fig. \ref{fig:4}a, we report an example from the failure of ultrahigh strength carbon fibers \cite{naito2012}. 
Experiments were performed at constant strain rate on a set of nominally identical fibers with equal volume. Hence,  it is possible to construct the survival curve  $\Sigma_V^{exp}(\{\sigma_i\})$ for each volume and strain rate and then fit the result using the rate dependent theory (Box 1). The resulting $\Lambda$ range from $10^{-3}$ to $10^{-2}$, 
indicating that we can safely neglect thermal and rate effects (case i). Indeed the survival curves all collapse into a
single curve when rescaled by the volume (see inset of Fig. \ref{fig:4}a)  despite the fact that strain rates
are different. In the second example reported in Fig. \ref{fig:4}b, CNT fibers are fractured at two very different
strain rates. In this case, however, volumes vary widely and it is not possible to construct a survival distribution
for each volume. Hence, the only available option  is to estimate the parameters by the maximum likelihood method.
The result indicates that $\Lambda > 1$ for the highest rate and we should therefore consider explicitly 
rate effects (case iv).

In general, the use of least-square fitting or maximum likelihood both present benefits and drawbacks \cite{wu2006,wu2006unbiased,ambrovzivc2007,ambrovzivc2011}. However, once the method is chosen, the problem is to select the correct statistical model for a given dataset, i.e. whether Weibull or Gumbel EVT distribution. In this respect, the maximum likelihood method is the most widely used to discriminate among competing models, by selecting the model with the largest likelihood function \cite{basu2009,cox1962}. 

\section*{Conclusions}

In this review article, we have discussed recent observations of size effects in the failure of materials at the micro and nanoscale. There is currently a growing interest in the development of smaller and smaller materials and devices with increasingly sophisticated experimental methods to test their mechanical properties. Understand and predict when nanomaterials will fail is of uttermost importance for all applications but the presence of size-dependent strength fluctuations makes the issue particularly complex. Reducing the sample scale typically make things worst: Small samples are more susceptible to thermal fluctuations and
even very small structural defects can have dramatic effects in driving failure. Even plastic deformation, that in macroscopic
samples is essentially size independent and smooth, becomes strongly size-dependent and intermittent at the micron scale and below.
The interpretation of these experimental observations needs novel theoretical tools that take disorder and fluctuations
explicitly into account.

Here we concentrated our discussion on EVT distributions that are currently widely used to fit experimental data obtained from micro and nanoscale samples. We highlighted that EVT are derived under well defined  assumptions and it is very important to understand if they are verified before embarking in a fit. The crucial assumption underlying EVT is that global failure is dictated by a single 
localized event, so that the theory does not necessarily hold when fracture arises from the coalescence of many localized events. Extensive simulations of disordered network models 
for fracture reveal that the fracture statistics is asymptotically ruled by EVT, even when
failure is preceded by diffuse damage accumulation \cite{Manzato2012,Bertalan2014,Shekhawat2013}. 
The convergence to EVT is shown to occur at relatively small scales \cite{Manzato2012,Shekhawat2013},
justifying the application of EVT to micro and even nanoscale materials. This is
confirmed by a wide variety of experimental and numerical results reviewed here. 

Despite the successes of EVT in describing a vast heterogeneity of experimental data, there are still cases where its application is more questionable. A notable example is the plastic deformation of crystalline materials, where yielding is due to the motion of a large number of interacting dislocations. The interesting issue of the origin of size effects and yield stress statistics in 
crystal plasticity is a topic of active research discussed in a number of excellent review articles
\cite{zaiser2006,kraft2010,greer2011}. 

Even when failure is ruled by extreme events, an important issue to be considered is the possible relevance of rate and thermal effects during the failure process. We have shown that it is possible to assess this by evaluating a simple dimensionless parameter. The second important issue stems from the nature of the data. In ideal conditions, one would need to compute the fracture strength of a large set of nominally identical samples, but this is rarely the case. In most cases, however, we have only access to a set of samples of varying volumes and possibly strain-rates. Even in this cases it is still possible to estimate useful parameters from the data using the maximum likelihood method. We hope the guidelines we propose here could be useful to understand size effects in micro and nanoscale materials and plan for future experiments in the field.

\clearpage

 \noindent {\bf BOX 1}: \\
{\it Extreme value theory}: According to the theory, the survival distribution function $\Sigma_V(\sigma)$, that is defined as the probability that a specimen of volume $V$ remains intact up to a stress $\sigma$, is given by 
\begin{equation}
\Sigma_V(\sigma)=\left\{ 
  \begin{array}{ccc}
   e^{-\left(\frac{\sigma}{\lambda}\right)^k} & & \sigma\in R^+ \mbox{~~Weibull} \\ 
   e^{-e^{\frac{\sigma-\mu}{\beta}}} & & \sigma\in R \mbox{~~Gumbel}
  \end{array}
\label{SigmaV}
  \right.
\end{equation}
where the volume dependence is hidden in the parameters, so that $\lambda=\lambda_0 (V/V_0)^{1/k}$ and 
$\mu/\beta=\mu_0/\beta_0 - \log V/V_0$

The probability density function (pdf) is given by $\rho_V(\sigma)=-\frac{d\Sigma_V(\sigma)}{d\sigma}$: 
\begin{equation}
\rho_V(\sigma)=\left\{
  \begin{array}{ccc}
\frac{k}{\lambda}\left(\frac{\sigma}{\lambda}\right)^{k-1}e^{-\left(\frac{\sigma}{\lambda}\right)^k} & & \sigma\in R^+\\ 
     \frac{1}{\beta}e^{\frac{\sigma-\mu}{\beta}-e^{\frac{\sigma-\mu}{\beta}}} & & \sigma\in R,
  \end{array}
  \right.
  \label{pdf_V}
\end{equation}
\noindent and size effects emerge  clearly after calculating the average strength 
\begin{equation}
\langle\sigma\rangle_V=\left\{
  \begin{array}{ccc}
    \lambda_0\left(\frac{V_0}{V}\right)^{1/k}\Gamma\left(1+\frac{1}{k}\right) & & \sigma\in R^+\\ 
     \mu_0+\ln\left(\frac{V_0}{V}\right)+\beta\gamma & & \sigma\in R
  \end{array}
\right.
\end{equation}
where $\Gamma$ represents the gamma function, and $\gamma$ is the Euler-Mascheroni constant.

\begin{figure}
  \centering
\includegraphics[width=10cm]{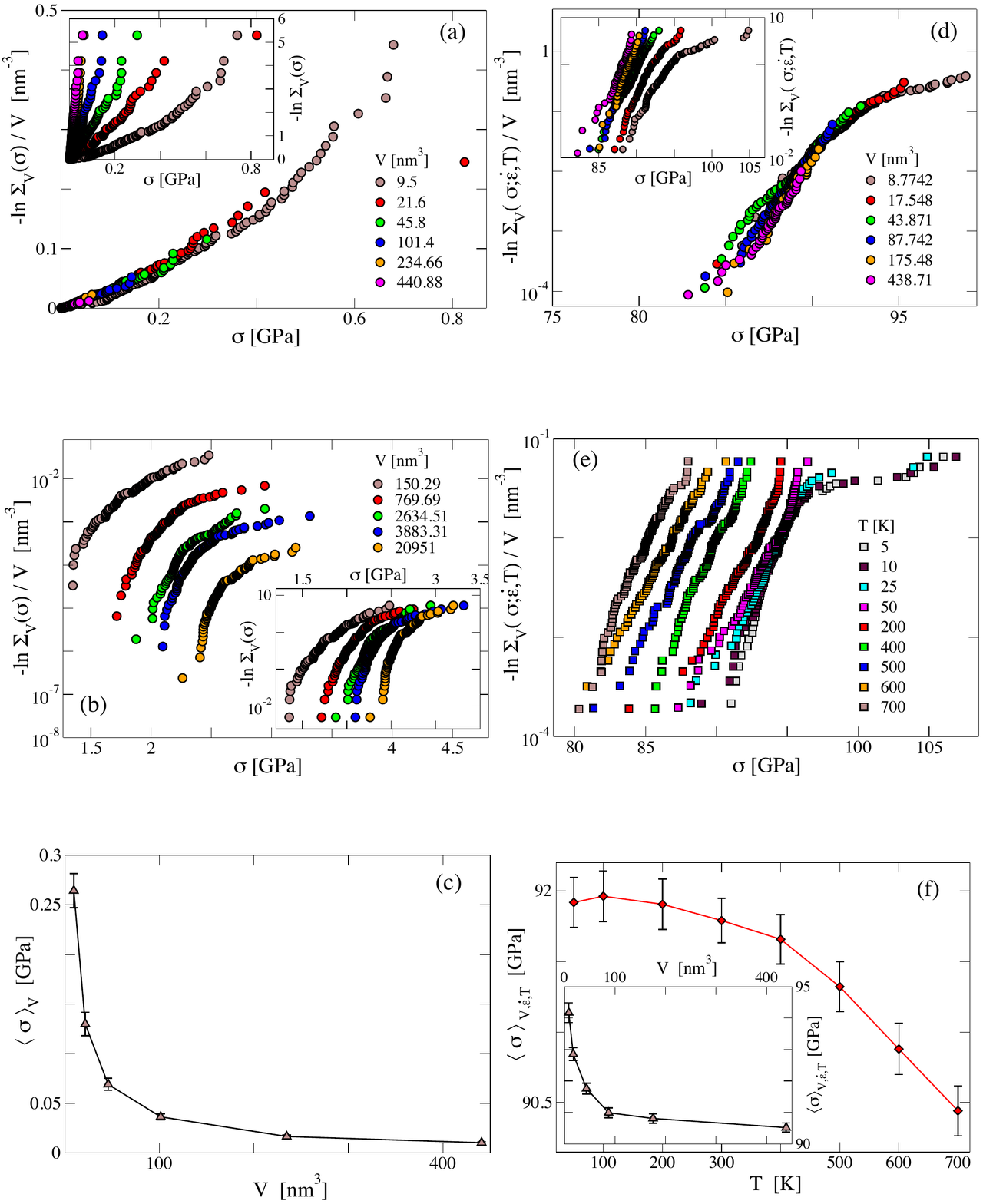}
\caption{{\bf Size dependence and EVT.} {\bf a}  Synthetic data for random failure strengths are generated according the Weibull distribution for several sample sizes $V$ (inset). Different $\Sigma_V$ are then rescaled, showing the collapse of the curves. This confirms that the hypothesis underlying EVT is fulfilled: the elementary volumes are independent. {\bf b} When the volume elements are not independent as for the correlated random variable plotted here, the survival distributions obtained for samples of  several volumes $V$ (inset) do not collapse when rescaled (main panel). {\bf c}  The average failure strength is calculated from the data shown in panel a. It displays a decrease for increasing sample size, according to the Weibull law.  {\bf d} When thermal and/or rate effects are non-negligible, the survival distribution can be constructed only from strengths obtained in experiments performed on equal volume samples $V$, and under the same experimental conditions $(\dot{\epsilon},T)$. The inset shows data coming from simulations of graphene sheets at different $V$, but same  $(\dot{\epsilon},T)$. Here data come from graphene MD simulations\cite{sellerio2015}. After rescaling of $\ln \Sigma_V(\sigma;\dot{\epsilon},T)$ by the sample volumes,  the curves show a nice collapse, 
confirming that the independence condition of the elementary volumes $V_0$ holds. {\bf e} $\Sigma_V(\sigma;\dot{\epsilon},T)$ arising from graphene sheets with same volumes $V$, same strain rates $\dot{\epsilon}$,  but different temperatures $T$. As the experimental conditions  among different datasets do not match, collapse of the survival distribution functions cannot be expected. However, as temperature decreases, thermal effects do not contribute to the failure tests leading to curve collapse (see curves for $T=5,10,25 K$). {\bf f}  Average failure strength as a function of $\dot{\epsilon}$ (inset) and $T$ (main panel). Data come from the same simulations in panels d and e.}
\label{fig:3}
\end{figure}

{\it Corrections due to thermal and rate effects.} When temperature and/or strain rate effects cannot be neglected, the weakest link hypothesis does not hold anymore: the first elementary volume $V_0$ which breaks does not necessarily coincide with the weakest, although its failure still causes the sample failure. As a consequence, the survival probability expression does not correspond to any of the EVT limit distributions and depends explicitly on temperature and strain-rate. If we assume 
that global failure is still dictated by the failure of first volume element, we can write the
survival distribution as  
\begin{equation}
\Sigma_V(\sigma;\dot{\varepsilon},T)=\left[\int_{\sigma/E}^{\infty} d\varepsilon_f \rho_0(\varepsilon_f)S_0(\sigma;\dot{\varepsilon},T|\varepsilon_f)\right]^{\frac{V}{V_0}}.
\label{Sigma_V_thermal}
\end{equation}
\noindent Here $\rho_0(\varepsilon_f)$ represents the failure strain distributions of the elementary volumes $V_0$,  assuming that they are perfectly brittle. The thermal factor $S_0$ can be derived from the Kramer's theory for the transition rate as \cite{sellerio2015}
\begin{equation}
S_0(\sigma;\dot{\varepsilon},T|\varepsilon_f)=e^{-\frac{\omega_0}{\sqrt{2\pi \dot{\varepsilon}}}\sqrt{\frac{EV_0}{k_BT}}\left[e^{-\frac{EV_0}{2k_BT}\left(\varepsilon_f-\sigma/E\right)^2}-e^{-\frac{EV_0}{2k_BT}\varepsilon_f^2}\right]},
\label{thermal_factor}
\end{equation}
\noindent where $\omega_0$ is a characteristic frequency, $E$ is the Young modulus and $k_B$ the Boltzmann's constant. 
Rate effects can be ignored when $\Lambda=\frac{\omega_0}{\dot{\varepsilon}} \sqrt{\frac{E V_0}{k_BT}} \ll 1$ and the thermal factor tends to 1,
recovering classical EVT (see \cite{sellerio2015} for a detailed derivation). According to  Eqs. \ref{Sigma_V_thermal}-\ref{thermal_factor}, the average failure stress decreases with volume $V$ and temperature (see Fig.\ref{fig:3}(f)) and increases with the strain rate $\dot{\varepsilon}$.

\noindent {\bf BOX 2}: \\
\emph{Case i: Samples of equal volume ($V_i\equiv V$), no rate and thermal effects}. In this case, it is useful to construct the experimental survival distribution function 
$\Sigma_V^{exp}(\{\sigma_i\})$, and to compare it with EVT distributions (Eq. \ref{SigmaV}). This is done in practice by  performing a linear regression of 
$\ln\left[-\ln \Sigma_V^{exp}(\{\sigma_i\})\right]$ with the fitting function $k\ln\sigma-k\ln\lambda$ for the Weibull distribution and  $\frac{\sigma}{\beta}-\frac{\mu}{\beta}$ for the Gumbel distribution. An equivalent alternative method  is to maximize the likelihood function $L(\{\theta\}|\{\sigma_i\})=\Pi_{i=1}^{n}\rho_V(\sigma_i,\{\theta\})$ where  $\rho_V(\sigma_i)$ is computed according to (\ref{pdf_V}) with parameters 
$\{\theta\}=(k,\lambda)$ for Weibull and $\{\theta\}=(\mu,\beta)$ for Gumbel.\\
\emph{Case ii: Samples of different volumes, no rate and thermal effects}.
 If each measured strength comes from samples with different volumes $V_i$, the maximum likelihood estimation method is the only viable strategy. Here, the the function takes the form
  $L(\{\theta\}|\{\sigma_i,V_i\})=\Pi_{i=1}^{n}\rho_{V_i}(\sigma_i)$, with 
$\{\theta\}\equiv (k,\alpha_0)$, where $\alpha_0=\lambda_0^kV_0$ for Weibull and  
$\{\theta\}\equiv (k,\nu_0)$ with $\nu_0=e^{\frac{\mu_0}{\beta}}V_0$ for Gumbel.\\
\emph{Case iii: Samples of equal volume, with rate and thermal effects}.
As in case i, when samples share the same volume we can compute the experimental survival distribution function $\Sigma_V^{exp}(\{\sigma_i\};\dot{\varepsilon},T)$ which will now
depend on strain rate and temperature. We can use the theory discussed in Box 1 (Eq. \ref{Sigma_V_thermal}) to perform least square fitting. Notice that the survival distributions should be computed separately for samples deformed  under different conditions of strain-rate and temperature. An alternative method relies in the maximum likelihood method.\\
\emph{Case iv: Samples of different volume, with rate and thermal effects}.
In this case, the only available method is the maximum likelihood estimation where the 
likelihood function is based on the rate and temperature dependent probability
density function as discussed in Box 1. 

\bibliography{fracture_refs-1}

\clearpage

\clearpage


%


	\clearpage

\setcounter{page}{1}

{\bf Supplementary Table S1.} Acronyms listed in the table: MWCNT = multiwalled carbon nanotube, AD = arc discharge, CVD = chemical vapor deposition, PAN =  poly-acrylonitrile, 
   AFM = atomic force microscope, AMCs = Al matrix composites, NW = nanowire, MDE = membrane deflection experiment, MG =  metallic glasses, \emph{od} = outer diameter, 
   \emph{l} = length, \emph{ar} = aspect ratio, \emph{t} = thickness, \emph{w} = width, \emph{uw} = upper width, \emph{lw} = lower width, \emph{sd} = short dimension, 
   \emph{ld} = long dimension, \emph{r} = radius.

  \begin{center}
  \begin{tabular}{|c|c|c|c|}
    \hline
    {\bf Material and size} &  {\bf Loading conditions} &  {\bf Statistical Analysis} &  {\bf Rate/Thermal effects}\\

    \hline
    \makecell{MWCNT (AD)~\cite{yu2000} \\ \emph{od} = 13-40 nm \\ \emph{l} = 1.10-10.99  $\mu$m}
    & tensile (AFM)
    & Weibull \cite{barber2005stochastic}
    & \makecell{none}\\

    \hline
    \makecell{MWCNT (CVD)~\cite{barber2005} \\ \emph{od} = 24-144 nm \\ \emph{l} = 10$\pm$4  $\mu$m}
    & tensile (AFM)
    & Weibull \cite{barber2005stochastic}
    & \makecell{none}\\

   \hline
   \makecell{MW WS$_2$ NT~\cite{kaplan2006} \\ \emph{od} = 11-36 nm \\ \emph{l} =0.85-2.95  $\mu$m}
   & tensile (AFM)
   & Weibull \cite{barber2005stochastic,kaplan2006}
   & \makecell{none}\\

   \hline
   \makecell{MWCNT fibers~\cite{sun2012} \\  \emph{d} = 4 $\mu$m \\ \emph{l} = 6  mm}
   & tensile \cite{zheng2010}
   & Weibull
   & \makecell{rate}\\

   \hline
   \makecell{CNT fibers (PAN)~\cite{zussman2005} \\  \emph{d}= 10–400 nm \\ \emph{ar} $\simeq$ 1000}
   & tensile (AFM)
   & Weibull
   & \makecell{none}\\

   \hline
   \makecell{CNT fibers~\cite{deng2011} \\ \emph{d}  = 20-50 $\mu$m}
   & tensile
   & Weibull
   & \makecell{none}\\



   \hline
   \makecell{C fibers (PAN/pitch) ~\cite{naito2012} \\  \emph{d} = 5.06, 5.22, \\ 7.37, 5.13,\\ 11.68, 9.35 $\mu$m \\ \emph{l} = 1, 5,\\ 12.5, 25,\\ 50, 100, 250 mm}
   & tensile
   & Weibull
   & \makecell{rate}\\

   \hline
   \makecell{C fibers (PAN) ~\cite{chae2015} \\  \emph{d} = 4.975 $\mu$m \\ \emph{l} = 1,  25.4  mm}
   & tensile
   & none
   & \makecell{none}\\

   \hline
   \makecell{MWCNT  (CVD) in AMCs~\cite{chen2015load} \\  \emph{d} = 1, 1.5 $\mu$m }
   & tensile
   & none
   & \makecell{none}\\

   \hline
   \makecell{Monolayer graphene flakes~\cite{Lee2008} \\  \emph{d} $\simeq$ 100 nm \\ \emph{t} $\simeq$ 1  nm}
   & indentation (AFM)
   & none
   & \makecell{none}\\

   \hline
   \makecell{Mono- and bi-layers \\of graphene films ~\cite{ruiz2011} \\  \emph{l} $\sim$ 3.5 $\mu$m}
   & indentation (AFM)
   & none
   & \makecell{none}\\

   \hline
   \makecell{SiC whiskers~\cite{petrovic1985}  \\  \emph{d} = 5-6 $\mu$m  \\ \emph{l} $\sim$ 10  mm }
   & tensile
   & Weibull
   & \makecell{none}\\

   \hline
   \makecell{SiC (curved, circular-hole,\\ elliptical-hole) ~\cite{nemeth2007}  \\ \emph{w} = 0.125mm  \\ \emph{l} =  1.3mm \\ \emph{t} = 0.2mm}
   & tensile
   & Weibull
   & \makecell{none}\\

   \hline
   \end{tabular}

   \begin{tabular}{|c|c|c|c|}
   \hline
   {\bf Material and size} &  {\bf Loading conditions} &  {\bf Statistical Analysis} &  {\bf Rate/Thermal effects}\\

   \hline
   \makecell{Si NW~\cite{fujii2013} \\ \emph{w} = 57-235 nm (FIB),\\149-314 nm (annealed) \\ \emph{t} =  221-444 nm (FIB),\\ 263-418 nm (annealed)\\ \emph{l} = 5.0  $\mu$m }
   & tensile (MEMS device)
   & none &
   \makecell{none}\\

   \hline
   \makecell{Si NW~\cite{tang2012mechanical} \\  \emph{d}  $\simeq$  25.3, 8.6 nm }
   & \makecell{tensile (TEM-AFM), \\bending}
   & none
   & \makecell{none}\\

   \hline
   \makecell{Si beams~\cite{namazu2000} \\ \emph{w} =  200, 300, 800 nm \\ \emph{t} =  255 nm\\ \emph{l} = 6.0  $\mu$m}
   & bending (AFM)
   & Weibull
   & \makecell{none}\\



   \hline
   \makecell{Si beams~\cite{li2003}  \\ \emph{uw} =  400 nm \\ \emph{lw} = 800 nm \\ \emph{l} = 6 $\mu$m }
   & bending (nanoindenter)
   & none
   & \makecell{none}\\

   \hline
   \makecell{Si beams~\cite{sundararajan2002}  \\ \emph{uw} =  200-600 nm \\ \emph{lw} = 385-785 nm\\ \emph{t} = 255 nm\\ \emph{l} = 6.0 $\mu$m }
   & bending(AFM)
   & Weibull
   & \makecell{none}\\

   \hline
   \makecell{SiO$_2$ beams~\cite{sundararajan2002}  \\ \emph{uw} =  250-700 nm \\ \emph{lw} =560-990 nm\\ \emph{t} = 425 nm\\ \emph{l} = 6.0 $\mu$m}
   & bending (AFM)
   & Weibull
   & \makecell{none}\\

   \hline
   \makecell{3C-SiC, UNCD, ta-C films~\cite{espinosa2006}  \\  \emph{sd} =  5, 20 $\mu$m \\ \emph{ld} = 100, 200 $\mu$m}
   & MDE
   & Weibull
   & \makecell{none}\\



   \hline
   \makecell{SCS  ~\cite{kang2013}  \\  \emph{w}=  720nm-8.7$\mu$m \\ \emph{l} = 6.5-8.7$\mu$m}
   & \makecell{thermomechanical\\ bending}
   & none
   & \makecell{temperature}\\

   \hline
   \makecell{SCS ~\cite{uesugi2015}  \\  \emph{w} =  4-9$\mu$m \\ \emph{t} = 5$\mu$m\\ \emph{l} = 120$\mu$m}
   & tensile  test
   & none
   & \makecell{temperature}\\

   \hline
   \makecell{Poly-Si ~\cite{boyce2010}  \\  \emph{w} =  3745, 3819, 3854 3922 nm \\ \emph{t} = 1029, 999, 958, 970 nm }
   & \makecell{tensile\\ (``pull-tab'' method \cite{boyce2007})}
   & none
   & \makecell{none}\\

   \hline
   \makecell{Poly-Si (central hole,\\ symmetric double notches) ~\cite{bagdahn2003} \\ \emph{w} =  20, 50 $\mu$m \\ \emph{t} = 3.5$\mu$m}
   & tensile
   & Weibull
   & \makecell{none}\\

   \hline
   \makecell{Poly-Si\\ (circular, elliptical hole) ~\cite{mccarty2007} \\  \emph{w}  =  30-340 $\mu$m \\ \emph{l} = 250-700 \\ \emph{t} = 2$\mu$m}
   & tensile 
   & Weibull
   & \makecell{none}\\

   \hline
   \makecell{Poly-Si films ~\cite{vayrette2015 } \\  \emph{w} =  1, 2, 4, 6, 8  10 $\mu$m  \\ \emph{t} = 40, 240 nm }
   & tensile 
   & none
   & \makecell{none}\\

   \hline
   \end{tabular}

   \begin{tabular}{|c|c|c|c|}
   \hline
   {\bf Material and size} &  {\bf Loading conditions} &  {\bf Statistical Analysis} &  {\bf Rate/Thermal effects}\\

   \hline
   \makecell{SiO$_2$ NW ~\cite{brambilla2009ultimate} \\  \emph{r} =  81, 94, 87, 96 nm  \\ \emph{l} $\simeq$ 6 mm }
   & tensile
   & none
   & \makecell{none}\\

   \hline
   \makecell{SiO$_2$ nanofibers ~\cite{luo2015size} \\  \emph{d} =  1-47.5 nm }
   & tensile 
   & none
   & \makecell{rate}\\

   \hline
   \makecell{Cu single crystal needles ~\cite{kiener2008}\\ \emph{d} =$0.5- 8 \mu$m\\ \emph{ar} = 1-13.5}    
   & tensile
   & none 
   & \makecell{none}\\

   \hline
   \makecell{Cu films ~\cite{lin2010} \\ \emph{w} = $100 \mu$m \\ \emph{l} =$500-600\mu$m}    
   & \makecell{tensile}
   & none 
   & \makecell{none}\\

   \hline
   \makecell{Cu single crystal pillars~\cite{jennings2011tensile}\\ \emph{d} =$75- 165$ nm, \\\emph{ar} = 3-6 }    
   & \makecell{tensile, \\ compression}
   & none
   & \makecell{none}\\

   \hline
   \makecell{Al ~\cite{haque2002}\\ \emph{w} =$23 \mu$ m \\ \emph{t}  = 200 nm \\ \emph{l} = 185 $\mu$m}    
   & tensile (MEMS device)
   & none
   & \makecell{none}\\



   \hline
   \makecell{Cu NW \cite{yue2013}\\ \emph{d} =5nm\\ \emph{l} = 45nm}    
   & tensile 
   & none 
   & \makecell{none}\\

   \hline
   \makecell{Cu beams ~\cite{motz2005} \\ \emph{w} =$2.5-7.5\mu$m,\\ \emph{t} =$1-7.5 \mu$m \\ \emph{l}=$20-25 \mu$m}    
   & bending
   & none
   & \makecell{none}\\

   \hline 
   \makecell{Cu beams ~\cite{kiener2006}\\  \emph{l} =$1-8\mu$m,\\ \emph{ar} = $1.5-2$}    
   & \makecell{ compression \\ bending}
   & none
   & \makecell{none}\\

   \hline
   \makecell{Cu films ~\cite{florando2005} \\ \emph{t} =$0.5,1, 1.7 \mu$m}    
   & bending 
   & none
   & \makecell{none}\\

   \hline
   \makecell{Cu films ~\cite{wei2007} \\ \emph{t} =$173-998$ nm}    
   & bulge \cite{vlassak1992}
   & none 
   & \makecell{none}\\

   \hline
   \makecell{Cu films ~\cite{merle2012}\\ \emph{t} =$400-2400$ nm}    
   & bulge \cite{vlassak1992}
   & none  
   & \makecell{none}\\

   \hline
   \makecell{Fe, Cu, Ag whiskers ~\cite{brenner1956}\\ \emph{d}  =$1.2-15 \mu$m}    
   &  tensile
   &  none 
   & \makecell{none }\\
   


   \hline
   \makecell{Cu whiskers ~\cite{richter2009}\\ \emph{d} =$20-300$ nm\\ l=$10-75 \mu$m}    
   &  tensile
   & none 
   & \makecell{none}\\

   \hline
   \makecell{Al films ~\cite{gianola2006} \\ \emph{t} =$180-380$ nm}    
   & tensile
   &  none 
   & \makecell{ none}\\

   \hline
   \makecell{Au beams ~\cite{gupta2017}\\ \emph{w} =$1.8\mu$m,\\ \emph{l}=$20 \mu$m \\ \emph{t}=$10$nm}    
   & tensile
   &  none 
   & \makecell{ none }\\

   \hline
   \makecell{Cu pillars~\cite{jennings2010microstructure}\\ \emph{d} =$100-500$ nm, \\ \emph{ar}= 3-6}    
   & \makecell{experimental compression}
   & none 
   & \makecell{none }\\

   \hline
   \makecell{Cu pillars~\cite{jennings2011}\\  \emph{d}=$75-500$ nm }    
   & compression
   & none 
   & \makecell{rate}\\

   \hline
   \makecell{W NW~\cite{wang2015}\\ \emph{d} =$ 14.7, 21$ nm }    
   & compression
   &  none
   & \makecell{ none}\\

   \hline
   \makecell{Mg pillars~\cite{lilleodden2010}\\ \emph{d} =$2.1, 6.1, 10 \mu$m }    
   & compression
   &  none
   & \makecell{ none}\\   

   \hline
   \end{tabular}

   \begin{tabular}{|c|c|c|c|}
   \hline
   {\bf Material and size} &  {\bf Loading conditions} &  {\bf Statistical Analysis} &  {\bf Rate/Thermal effects}\\

   \hline
   \makecell{Al pillars~\cite{kunz2011}\\ \emph{d} =$400$ nm-2 $\mu$m, \\ \emph{ar} = 3-4}    
   & compression
   & none
   & \makecell{none}\\

   \hline
   \makecell{W, Mo, Ta, Nb pillars~\cite{kim2010}\\ \emph{d} =$100-900$ nm}    
   & tensile, compression
   & none
   & \makecell{ none}\\

   \hline
   \makecell{Au NW~\cite{wang2013} \\ \emph{d} =$8-20$ nm}    
   & tensile
   & none
   & \makecell{ none}\\  

   \hline
   \makecell{Cu films~\cite{sim2013}\\ \emph{w} =$100, 150, 200 \mu$m, \\ \emph{l}=$1326, 2552 \mu$m\\ \emph{t}=$385, 465, 800$ nm}    
   & tensile
   & none
   & \makecell{temperature}\\

   \hline
   \makecell{Au films~\cite{sim2014}\\ \emph{w} =$100, 150, 200 \mu$m, \\ \emph{l}=$1232, 2425 \mu$ m\\ \emph{t}=$450, 960$ nm}    
   & tensile
   & none
   & \makecell{rate}\\

   \hline 
   \makecell{poly-Si, Cu films ~\cite{vayrette2015}\\ \emph{t} =$240-290$ nm}    
   & tensile
   & none
   & \makecell{none}\\



   \hline
   \makecell{Zr-based MG pillars~\cite{lai2008bulk}\\ \emph{d}=$3.8, 1, 0.7  \mu$m}    
   & compression
   & \makecell{Weibull}
   & \makecell{rate}\\

  \hline   
  \makecell{Pd-based MG pillars~\cite{schuster2007bulk}\\ \emph{d}=$2-10 \mu$ m\\ \emph{ar} = 2-2.5}    
  & compression
  & \makecell{Weibull}
  & \makecell{none }\\ 

  \hline 
  \makecell{Zr-based MG pillars~\cite{chen2015microstructural}\\ \emph{d}=$200$ nm}    
  & compression
  & \makecell{Weibull}
  & \makecell{rate}\\

  \hline 
  \makecell{Co-based ribbons~\cite{calvo1989}\\ \emph{w} = 3.7-4.9 mm \\  \emph{t} =m 47-57 $\mu$m\\ \emph{l} = 50 mm}
  & tensile
  & \makecell{Weibull}
  & \makecell{rate}\\

 \hline 
  \makecell{Ni$_{80}$Si$_{80}$B$_{10}$ and Ni$_{80}$Si$_{5}$B$_{15}$ ribbons~\cite{ocelik1987}}
  & tensile
  & \makecell{Weibull}
  & \makecell{rate, thermal}\\

  \hline 
  \makecell{Mg-based pillars~\cite{ocelik1987}\\ \emph{d} = 1.86 mm\\ \emph{l} = 3.5-4 mm}
  & compression
  & \makecell{Weibull}
  & \makecell{none}\\

  \hline 
  \makecell{Zr$_{48}$Cu$_{45}$Al$_{7}$ and
(Zr$_{48}$Cu$_{45}$Al$_{7}$)$_{98}$Y$_2$~\cite{wu2008}\\ \emph{d} = 1.5 mm\\ \emph{l} = 3.1-3.3 mm}
  & compression
  & \makecell{Weibull}
  & \makecell{none}\\ 

  \hline 
  \makecell{Zr-based rods ~\cite{han2009}\\ \emph{d} = 1.5 mm\\ \emph{l} = 30 mm}
  & compression
  & \makecell{Weibull}
  & \makecell{none}\\

  \hline 
  \makecell{Fe-based rods ~\cite{shamimi2008}\\ \emph{d} = 1.5-4 mm\\ \emph{l} = 30-40 mm}
  & bending
  & \makecell{Weibull}
  & \makecell{none}\\

  \hline 
  \makecell{Zr$_{48}$Cu$_{45}$Al$_{7}$ ~\cite{yao2008}\\ \emph{w} = 1 mm\\ \emph{t} = 0.7mm \emph{l} = 4 mm}
  & tensile
  & \makecell{Weibull}
  & \makecell{none}\\

  \hline 
  \makecell{Zr$_{63.8}$Ni$_{16.2}$Cu$_{15}$Al$_{5}$,\\ Pd$_{40}$Ni$_{40}$
    P$_{20}$,\\ Au$_{49}$Ag$_{5.5}$Pd$_{2.3}$Cu$_{26.9}$Si$_{16.34}$,\\ Mg$_{65}$Cu$_{25}$Gd$_{10}$ pillars ~\cite{lee2010}\\ \emph{d} = 3.8 $\mu$m\\ \emph{l} = 9 $\mu$m}
  & compression
  & \makecell{Weibull}
  & \makecell{none}\\

  \hline 
  \makecell{Zr$_{35}$Ti$_{30}$Co$_{6}$Be$_{29}$  specimens ~\cite{jang2010transition}\\ \emph{d} = 330-949 nm\\ \emph{l} = 3-5 $\mu$m, 800 nm}
  & tension
  & \makecell{none}
  & \makecell{none}\\

  \hline 
  \makecell{Zr$_{35}$Ti$_{30}$Co$_{6}$Be$_{29}$  specimens ~\cite{jang2010transition}\\ \emph{d} = 330-949 nm\\ \emph{l} = 3-5 $\mu$m, 800 nm}
  & tension
  & \makecell{none}
  & \makecell{none}\\

  \hline 
  \makecell{Zr$_{35}$Ti$_{30}$Co$_{6}$Be$_{29}$  pillars ~\cite{jang2011effects}\\ \emph{d} = 117-1710 nm\\ \emph{l} = 0.9-7.37 $\mu$m}
  & compression
  & \makecell{Weibull}
  & \makecell{none}\\

\hline
  \end{tabular}

  \end{center}
  
\end{document}